\documentstyle[12pt]{article}

\newcommand{\la}{\label}
\newcommand{\Pfaff}{{\rm\, Pfaff}}

\newcommand{\E}{{\cal E}}

\newcommand{\X}{{\cal X}}
\renewcommand{\O}{{\cal O}}
\renewcommand{\d}{{\rm\, d}}
\newcommand{\e}{{\rm\, e}}
\newcommand{\non}{\nonumber}
\newcommand{\be}{\begin{equation}}
\newcommand{\ee}{\end{equation}}
\newcommand{\ba}{\begin{eqnarray}}
\newcommand{\ea}{\end{eqnarray}}
\newcommand{\half}{{1 \over 2}}
\begin{document}
\begin{titlepage}
\hfill hep-th/9508067
\vskip 0.7truecm

\begin{center}
{ \bf TOPOLOGICAL $\sigma$-MODEL, HAMILTONIAN DYNAMICS  \\ \vskip 0.3cm
AND LOOP SPACE LEFSCHETZ NUMBER
      }
\end{center}

\vskip 1.5cm

\begin{center}
{\bf Antti J. Niemi $^{*\dagger}$ } \\

\vskip 0.3cm

{\it Department of Theoretical Physics, Uppsala University \\
P.O. Box 803, S-75108, Uppsala, Sweden $^{\ddagger}$ \\

\vskip 0.2cm

and \\

\vskip 0.2cm

Research Institute for Theoretical Physics \\
P.O.Box 9,  FIN-00014 University of Helsinki, Finland} \\

\vskip 0.5cm

{\bf Pirjo Pasanen $^{**}$ } \\
\vskip 0.3cm
{\it Research Institute for Theoretical Physics \\
P.O. Box 9, FIN-00014 University of Helsinki, Finland}

\end{center}

\vskip 1.5cm
\rm
\noindent
We use path integral methods and topological quantum field theory
techniques to investigate a generic classical Hamiltonian system.  In
particular, we show that Floer's instanton equation is related to a
functional Euler character in the quantum cohomology defined by the
topological nonlinear $\sigma$--model.  This relation is an infinite
dimensional analog of the relation between Poincar\'e--Hopf and
Gauss--Bonnet--Chern formul\ae$~$ in classical Morse theory, and can
also be viewed as a loop space generalization of the Lefschetz fixed
point theorem.  By applying localization techniques to path integrals
we then show that for a K\"ahler manifold our functional Euler
character coincides with the Euler character determined by the finite
dimensional de Rham cohomology of the phase space.  Our results are
consistent with the Arnold conjecture which estimates periodic
solutions to classical Hamilton's equations in terms of de Rham
cohomology of the phase space.

\vfill

\begin{flushleft}
\rule{5.1 in}{.007 in} \\
$^{\ddagger}$  \small permanent address \\
$^{\dagger}$ \small Supported by G{\"o}ran Gustafsson Foundation
for Science and
Medicine \\
\hskip 0.3cm and by NFR Grant F-AA/FU 06821-308  \\ \vskip 0.2cm
$^{*}$ \hskip 0.15cm \small E-mail: $~~$ \small
NIEMI@TETHIS.TEORFYS.UU.SE  \\
$^{**}$ \small E-mail: $~~$ \small PIRJO.PASANEN@HELSINKI.FI \\
\end{flushleft}

\end{titlepage}
\vfill\eject

The methods of quantum field theory were originally developed to
understand particle physics, but have since proven useful also in
statistical physics. Recently there have been various indications
that these methods might be successfully applied even in the study of
{\it classical} Hamiltonian dynamics. One of the intriguing
open problems there is the Arnold conjecture \cite{Arnold}, 
\cite{Hofer} which states that on a compact phase space the 
number of periodic solutions to Hamilton's equations is bounded 
from below by the sum of Betti numbers on the phase space.

In the present Letter we shall argue that topological quantum
field theories \cite{blau} and path integral localization techniques
\cite{omat} are indeed potentially effective tools in the study of
classical dynamical systems.  Furthermore, by applying
such methods to investigate classical structures one might also
gain valuable insight to the nonperturbative structure of quantum
dynamics.

We shall consider Hamilton's equations on a phase space which is a
compact symplectic manifold $X$ with local coordinates $\phi^a$. We
are interested in $T$-periodic trajectories that solve Hamilton's
equations, {\it i.e.} are critical points of the classical action
\be
S_{\rm cl} = \int_0^T \d\tau (
\vartheta_a\dot\phi^a - H(\phi,\tau) )\la{clasact}
\ee
Here $\vartheta_a$ are components of the symplectic potential
corresponding to the symplectic two-form $\omega =\d\vartheta$.
 We assume that the
Hamiltonian depends {\it explicitly} on time $\tau$ in a $T$-periodic
manner $H(\phi , 0) = H(\phi , T)$, so that energy is not necessarily
conserved.  Hamilton's equations are
 \be
\dot\phi^a -
\omega^{ab}\partial_b H(\phi;\tau) \ = \ \dot\phi^a -\X_H^a =0
\la{trajectory}
\ee
with $T$-periodic boundary condition $\phi(0) =
\phi(T)$. We are particularly interested in the case
where the periodic solutions are nondegenerate.

When energy is conserved so that $H$ does not have explicit dependence
on $\tau$ each critical point $\d H(\phi)=0$ of $H$ generates trivially a
$T$-periodic trajectory. According to the classical Morse theory the
number of these critical points
is bounded from below by the sum of Betti numbers on $X$ and
consequently Arnold's conjecture is valid:
\be
\#\{ \  T {\rm -periodic ~ trajectories} \ \}  ~ \geq ~  
\sum B_k = \sum {\rm
dim\,}H^k (X, R)
\la{Arconj}
\ee
But if $H$ depends explicitly on time so that energy is {\it not}
conserved, the critical points of $H$ do not solve (\ref{trajectory})
and the methods of finite dimensional Morse theory are no longer
applicable.  Instead, we need to estimate directly the quantity
\be
\#\{ \  T {\rm -periodic ~ trajectories} \ \}  ~ = ~
\int\limits_{\phi(T) = \phi(0)} [d\phi]
\delta(\dot \phi^a - \X^a_H ) \left| \det ||
\delta_a^b \partial_t + \partial_a \X^b_H || \right|
\la{estimate1}
\ee
and for this we need an
infinite dimensional generalization of Morse theory.  Unfortunately
this is not very easy. There is no minimum for (\ref{clasact}), and
periodic solutions of (\ref{trajectory}) are saddle points of
(\ref{clasact}) with an infinite Morse index\footnote{In the following
we shall define all determinants using $\zeta$-function
regularization.}. 
Due to such
difficulties, for explicitly $\tau$-dependent Hamiltonians the
conjecture has only been proven in certain special cases \cite{Hofer}.

An important special case is governed by the Lefschetz fixed point
theorem \cite{eguchi}: We first estimate the integral in
(\ref{estimate1}) by
\be
\geq ~ \left| \ 
\int\limits_{\phi(T) = \phi(0)} [d\phi]
\delta(\dot \phi^a - \X^a_H )  \det ||
\delta_a^b \partial_t + \partial_a \X^b_H || \right|
\ = \ \left| \sum\limits_{\delta S_{PBC} = 0} sign \left\{
\det || \delta_a^b \partial_t + \partial_a \X^b_H || \right\}
\right|
\la{estim2}
\ee
where on the {\it r.h.s.} we identify an infinite dimensional version
of the sum that appears in finite dimensional Poincar\'e-Hopf
theorem with $S$
as a Morse function. If we introduce a commuting variable $p_a$
and two anticommuting variables $c^a$, $\bar c_a$, we can
write the integral in (\ref{estim2}) as 
\be
\int [d\phi][dp][dc][d\bar c] \exp \{ i \int\limits_o^T 
p_a \dot\phi^a - \bar c_a \dot c^a - p_a \X^a_H - \bar c_a
\partial_b \X^a_H c^b \}
\la{estim3}
\ee
Furthermore, if we define the following nilpotent BRST operator
\be
Q \ = \ c^a {\partial \over \partial \phi^a } \ + \ p_a {\partial
\over \partial \bar c_a}
\la{estim4}
\ee
we find that this integral can be represented in the topological
form
\be
= ~ \int [d\phi][dp][dc][d\bar c] \exp \{ \  i \int\limits_0^T
p_a \dot\phi^a - \bar c_a \dot c^a - Q ( \X^a_H \bar c_a ) \ \}
\la{estim5}
\ee
According to standard BRST arguments, this integral depends
only on the cohomology of the BRST operator $Q$.
In particular we can replace $Q$ by its canonical conjugation
$Q \to e^{-\theta} Q e^{\theta}$ where we select
$\theta = -\Gamma^c_{ab} c^a \bar c_c \pi^b$, with 
$\Gamma^c_{ab}$ the connection for some metric
$g_{ab}$ on the phase space $X$ and $\pi^a p_b =
\delta^a_b$. Furthermore, if we generalize
\[
\X^a_H \bar c_a \ \to \ (1-\lambda) \X^a_H \bar c_a  \ + \lambda
g^{ab} p_a
\bar c_b
\] 
where $\lambda \in [0,1]$ is a parameter, 
BRST invariance implies that the ensuing
path integral (\ref{estim5}) is
independent of $\lambda$. Selecting $\lambda = 1$ we then
conclude that (\ref{estim5})
localizes to the Euler characteristic of the phase space $X$
\cite{blau}, \cite{oma2}. Consequently
\be
\#\{ \  T {\rm -periodic ~ trajectories} \ \} \ \geq \
\left| \int_X {\rm Pf} [ \half {R^a}_{bcd} c^c c^d ] \right|
\ = \ \left| \sum_k (-)^k H^k (X,R) \right|
\la{lefs}
\ee
and for manifolds with vanishing odd Betti numbers {\it i.e.}
$H_{2k+1}= 0$ this establishes the validity of Arnold's conjecture.
This result can be viewed as a path integral proof of the
Lefschetz fixed point theorem \cite{eguchi}: If $F:X\to X$ is a smooth
map which is homotopic to the identity and admits only isolated fixed
points $F[\phi]=\phi$, according to this theorem the number of these
fixed points is bounded from below by the Euler character of $X$. In
the present context, Hamilton's equations determine such a smooth map
$F: X \to X$ by $F[\phi] \equiv F[\phi(t=0)] \to \phi(T)$ and the time
evolution determined by Hamilton's equations ensures that this map is
homotopic to the identity.  The fixed point condition $F[\phi] = \phi$
then selects $T$-periodic classical trajectories $\phi(0)=\phi(T)$.

In spite of very extensive work \cite{Hofer} in the general case when
the odd Betti-numbers are non-vanishing the Arnold conjecture remains
unproven. However, recent attempts have led to  interesting
approaches, including topological nonlinear $\sigma$-models. 
In the approach originating from Floer \cite{Floer,Hofer} one starts by
defining a gradient flow in the space of closed loops $\phi(0) =
\phi(T)$
\be
\frac{\partial\phi^a}{\partial\sigma} =  - g^{ab}
\frac{\delta S_{\rm cl}}{\delta \phi^b}
\la{gradient}
\ee
where $g_{ab}$ is a Riemannian metric on $X$.  
Using this metric and the
symplectic two-form $\omega_{ab}$ we set
\[
{I^a}_b = g^{ac}\omega_{cb}
\]
Since ${I^a}_c{I^c}_b = -{\delta^a}_b$ this defines an  almost complex
structure on  the manifold $X$ and (\ref{gradient}) becomes
\be
\partial_\sigma \phi^a + {I^a}_b\partial_\tau\phi^b = \partial^a H
(\phi,\tau).
\la{pseudo}
\ee
This is Floer's instanton equation, defined on a cylinder $S^1\times
R$ with local coordinates $\tau\in [0,T]$ and $\sigma \in (-\infty,
\infty)$.  It describes the flow of loops $\phi(\tau)$ on $X$, and the
bounded orbits in $\sigma$ tend asymptotically to the periodic
solutions of Hamilton's equation (\ref{trajectory}).  Using
(\ref{pseudo}) Floer constructs a complex with the solutions to
(\ref{trajectory}) being the vertices and the trajectories
(\ref{pseudo}), so-called pseudo-holomorphic instantons, connecting
them as the edges. He proves the important result that the cohomology
of this complex is {\it independent} of the Hamiltonian
$H(\phi,\tau)$.  Subsequently Witten \cite{Wit} found that this Floer
cohomology can actually be related to a quantum cohomology which is
generated by the quantum ground states of a topological
$\sigma$--model.  Using the more general Novikov ring structure Sadov
\cite{Sadov} then argued that these two cohomologies in fact
coincide.

Witten's topological nonlinear 
$\sigma$-model \cite{Wit} is based on solutions
of Cauchy-Riemann equations for holomorphic curves
\be
\partial_\sigma \phi^a +
{I^a}_b\partial_\tau\phi^b =0
\la{CR}
\ee
His approach has led to interesting developments in quantum field
theories and string theory \cite{blau,CMR}, but
unfortunately from the point of view of Hamiltonian dynamics it
corresponds to the non-generic, denegerate special case of
(\ref{pseudo}) with $H=0$.  Consequently it is not clear how the
topological $\sigma$--model, even if it describes Floer's cohomology,
could be applied to understand Arnold's conjecture. For this, one
needs to extend the topological $\sigma$-model so that it accounts for
a {\it generic} nontrivial Hamiltonian $H(\phi, \tau)$, for which the
solutions to (\ref{trajectory}) are non-degenerate.

In the present Letter we shall explain how path integrals and
localization techniques, when applied to the topological
$\sigma$--model, can be used to derive Morse--theoretic relations for
classical trajectories in a generic Hamiltonian system. In particular,
we shall explain how the standard, finite dimensional de Rham
cohomology relates to quantum cohomology by studying an infinite
dimensional version of Poincar\'e--Hopf and Gauss--Bonnet--Chern
formul\ae$~$ for (\ref{pseudo}), and by an exact evaluation of the
partition function of the topological $\sigma$-model we also derive an
extension of the Lefschetz fixed point theorem for Floer's instanton
equation (\ref{gradient}), (\ref{pseudo}) on the cylinder.

\vskip 0.5cm

Topological nonlinear $\sigma$--model \cite{Wit} is a theory of maps
from a Riemann surface $\Sigma$ with metric $\eta_{\alpha\beta}$ and
almost complex structure ${\epsilon_{\alpha}}^\beta$ to a manifold $X$
with Riemannian metric $g_{ab}$ and almost complex structure
${I^a}_b$. We assume that the almost complex structures are both
compatible with the metrics, so that for example on $X$ we have
$g_{ab}= {I^c}_a{I^d}_bg_{cd}$. Moreover, if
\be
D_c{I^a}_b = \partial_c{I^a}_b +
\Gamma^a_{cd} {I^d}_b - \Gamma^d_{cb}{I^a}_d=0
\la{ctl}
\ee
${I^a}_b$ is an integrable complex structure and $g_{ab}$ is K\"ahler.
However, in the following we do not necessarily assume (\ref{ctl}).

The  fields are the space of 
maps $\phi^a : \Sigma \to X,\ a = 1 \ ... \ {\rm
dim\ } X$. Anticommuting
fields $\psi^a$ are sections of $\phi^* TX$, the pullback of the
tangent bundle of $X$. Anticommuting fields $\rho_\alpha^a,\, (\alpha
=1,2)$, and commuting
auxiliary fields $F_\alpha^a$ are one-forms on $\Sigma$ with values on
$\phi^* TX$, so they
are sections of the bundle $\phi^*TX \otimes T^*\Sigma$.
 Let $\E$ denote a bundle over the space of maps from
$\Sigma$ to $X$, whose fibers are sections of $\phi^*TX \otimes
T^*\Sigma$. Because the rank of $\E$ is infinitely bigger than
dimension of its base space we must restrict to a
sub-bundle, the self--dual part $\E^{+}$.  This means that
$\rho_\alpha^a$ and $F_\alpha^a$ both satisfy the self--duality
constraint
\be
\rho_\alpha^a = {\epsilon_\alpha}^\beta {I^a}_b \rho_\beta^b
\la{duality}
\ee
The fields have a grading, which at the classical level corresponds to a
bosonic symmetry with charges $0,1,-1,0$ for
$\phi^a,\psi^a,\rho^a_\alpha$ and $F^a_\alpha$, respectively.

The action of topological $\sigma$--model can be constructed in
the following way: Consider a nilpotent
operator $\tilde Q$ of degree
$-1$ constructed from the fields of the theory
\be
\tilde Q = \int_\Sigma d^2 x \left[ i\psi^a (x)
\frac{\delta}{\delta \phi^a(x)} +
F^a_\alpha (x)\frac{\delta}{\delta \rho^a_\alpha(x)} \right]
\equiv i\psi^a \partial_a + F^a_\alpha \iota^\alpha_a ,
\ee
(In the following summation over $a$ always implies an integration
over $\Sigma$.) This we identify as a differential operator $\d\otimes
1 +1 \otimes\delta$ in the superspace
defined on the complex $\Omega (\E)\otimes\Omega (\Pi\E)$.  Here
$\Pi\E$ means that the coordinates anticommute.  Now introduce a
canonical conjugation $\tilde Q \to e^{-\theta} Q e^\theta$ so that
the cohomologies defined by the operators $\tilde Q$ and $Q$ are the
same. A suitable conjugation is defined by
\[
\theta = i \psi^c \rho^b_\beta \pi^\alpha_a ({\delta_\alpha}^\beta
\Gamma^a_{bc} - \half {\epsilon_\alpha}^\beta D_c {I^a}_b ),
\]
where now $\pi_a^\alpha F^b_\beta = \delta_{a\beta}^{b\alpha}$.
In a coordinate free language $\theta = -i (\rho, \hat
\Gamma \pi)$, with
\[
\hat \Gamma ^{a\beta}_{\alpha b} =
{{\hat\Gamma^a}}_b\otimes {E_\alpha}^\beta =
{\delta_\alpha}^\beta
{\Gamma^a}_{b} - \half {\epsilon_\alpha}^\beta D {I^a}_b
\]
a connection 1-form
and $\psi^a \sim d\phi^a$ denoting the basis of 1-forms on the space
of maps $\Sigma \rightarrow X$.
A straightforward calculation gives
\[
Q= i (\psi , \partial )  + (F + i\rho \hat\Gamma , \iota)
-i(F\hat\Gamma , \pi) -\half (\rho \hat R, \pi).
\]
where
\[
\half \hat R \ = \ d\hat\Gamma + \hat\Gamma \wedge \hat\Gamma
\]
or in components
\be
\frac{1}{2} {{\hat R}^{a \alpha}}_{\beta b}~ = ~ (  \frac{1}{2}
{R^a}_{b} - \frac{1}{4} D{I^e}_b D{I^a}_e  ) {\delta_\beta}^\alpha
\ + \ \frac{1}{4} ( \  {I^a}_e {R^e}_{b} -  {I^e}_b {R^a}_{e} \ )
{\epsilon_\beta}^\alpha       .
\la{Riemann}
\ee
which is the Riemann curvature 2-form corresponding to the connection
$\hat
\Gamma^{a\beta}_{\alpha b}$. 
This operator $Q$ is exactly the same as in
\cite{Wit} when we take into account the self-duality 
condition (\ref{duality}) for $\rho^a_\alpha$ and $F^a_\alpha$.

\vskip 0.3cm

We shall be interested in cohomological actions of the form
\be
S = \{Q, \theta\}
\la{action}
\ee
Such actions are automatically invariant under the
BRST-transformation generated by $Q$
and consequently the partition function
\be
Z = \int [\d \phi^a] [\d F^a_\alpha ][\d 
\psi^a][\rho^a_\alpha] \e ^{iS}
\la{path}
\ee
should remain invariant under arbitrary local variations of  $\theta$.
If we select
\be
\theta = (\rho, s) - \frac{\lambda}{4}(\rho, F) ~=~ \rho^a_\alpha
g_{ab}\eta^{\alpha\beta}s^b_\beta -  \frac{\lambda}{4}  \rho^a_\alpha
g_{ab}\eta^{\alpha\beta}F^b_\beta , \non
\ee
where $s^a_\alpha [\phi] $ is a 
section of ${\cal E}$ and $\lambda$ is a
parameter, we get
\ba
S &=& \int_\Sigma \left[ -i\rho^a_\alpha D_c (g_{ab}s^{\alpha b} )
\psi^c + F_\alpha^a g_{ab} s^{\alpha b}
-\frac{i}{2} \epsilon_\alpha^\beta \rho_\beta^b D_c {I^a}_b \psi^c
g_{ad} s^{\alpha d} - \frac{\lambda}{4}
F_\alpha^a F^\alpha_a  \right.  \non\\
&+& \left. \frac{\lambda}{16} D_c {I^a}_e D_d {I^e}_b
\psi^c\psi^d \rho^\alpha_a \rho_\alpha^b  - \frac{\lambda}{8} {R^a}_{bcd}
\psi^c\psi^d \rho^\alpha_a\rho^b_\alpha \right] .
\la{Waction}
\ea
specializing to $s^a_\alpha[\phi] = 
\partial_\alpha \phi^a$ and $\lambda = 1$ then 
gives the usual action \cite{Wit} of topological $\sigma$--model.

Since the partition function (\ref{path}) is 
(formally) invariant under local
variations of $\theta$ we conclude that  it must be independent of
$\lambda$. Indeed, if we eliminate 
the auxiliary field $F^a_\alpha$, the
partition function yields an infinite
dimensional version of the Mathai--Quillen
formalism \cite{MQ,CMR}: 
\ba
S &=& \int_\Sigma \left[
\frac{1}{4\lambda} (s_\alpha^a +
{\epsilon_\alpha}^\beta {I^a}_b s^b_\beta)(s^\alpha_a +
{\epsilon^\alpha}_\beta {I_a}^b s_b^\beta)
-\frac{i}{2}\rho_a^\alpha D_c (s_\alpha^a +
{\epsilon_\alpha}^\beta {I^a}_b s^b_\beta)
\psi^c \right.
\non\\
&-&\left.
\frac{\lambda}{4} \hat {R^a}_{bcd} \psi^c\psi^d 
\rho^\alpha_a\rho^b_\alpha
\right]
\la{MQaction}
\ea
the relevant bundle being ${\cal E}^+$
and the section
\be
\Phi^a_\alpha ~=~ s^a_\alpha 
+ {\epsilon_\alpha}^\beta {I^a}_b s^b_\beta
\la{Phi}
\ee
Thus we may view (\ref{path}) as an infinite dimensional version of
the integral of the universal Thom class \cite{CMR}.

Since (\ref{path}) is independent of $\lambda$, we can consider its
$\lambda \to \infty$ limit.  For this, we specialize the world-sheet
$\Sigma$ to be a torus with $\sigma$ and $\tau$ local coordinates such
that the metric $\eta_{\alpha\beta}$ is a unit matrix with compatible
complex structure ${\epsilon_\sigma}^\tau = - {\epsilon_\tau}^\sigma =
1$.  In the $\lambda
\to \infty$ limit we then find 
that the partition function evaluates to
\be
Z_{\lambda\to\infty} ~ = ~
\int [\d\phi^a][\d\psi^a] \Pfaff (\hat {R^a}_b)
\la{euler}.
\ee
This we identify as the Euler character of the infinite dimensional
bundle $\E^{+}$.  Formally, this infinite dimensional quantity is a
topological invariant and as such does not depend on how we choose the
connection. It is the Euler character in the quantum cohomology
defined by the quantum ground states of the topological
$\sigma$--model, and by construction it
counts the Witten index {\it i.e.} the
difference in the number of bosonic vacua (even forms) and fermionic
vacua (odd forms) in the quantum theory.

In analogy with finite dimensional Morse theory, we next relate the
infinite dimensional Euler character (\ref{euler}) to an
alternating sum over critical points of a functional $\Phi$ describing
the Floer cohomology. For this we consider the limit $\lambda\to 0$,
again on a torus $\Sigma$ with local coordinates $\sigma, \tau$.

As $\lambda\to 0$, the integral obviously concentrates around the
zeroes of (\ref{Phi}).  For simplicity we shall assume that these
zeroes are non-degenerate. (A generalization to the degenerate case is
straightforward, see for example
\cite{NiePa}.) Let $\phi^a_0$ be such that
$\Phi^a_\alpha [\phi_0] =0$ and write $\phi^a = \phi_0^a +
\hat\phi^a$. In the absence of degeneracies, the first term in the
expansion
\[
\Phi_\alpha^a \approx \partial_c
(\Phi^a_\alpha)\hat\phi^c + \O (\hat\phi^2)
\]
does not vanish. Using the self-duality of $\rho^a_\tau$ 
and the fact that near
$\phi_0$ we have $\Phi^a_\tau = - {I^a}_b \Phi^b_\sigma$ 
this gives for the action
\be
S = \int_\tau \left[ -i\rho^a_\sigma g_{ab} \partial_c
(\Phi_\sigma^a )\psi^c
+ \frac{1}{2\lambda}
\partial_c(\Phi_\sigma^a )g_{ab} \partial_d(\Phi_\sigma^b )
\hat\phi^c \hat\phi^d + \O (\hat\phi^3)
\right].
\ee
As $\lambda \to 0$, we can then evaluate the partition function which
yields
\ba
Z_{\lambda\to 0}&=& \int [\d \phi_0^a][\d\hat\phi^a][\d \psi^a]
[\rho^a_\tau][\rho^a_\sigma] \det{}^{-\half}(\frac{i\lambda}{2}g)
\exp [iS] \non\\
&=&
\int [\d \phi_0^a]\det{}^{-\half}(g) \det{}^{-\half}\left[
 \left(\partial_c(\Phi_\sigma^a )g_{ab}
\partial_d(\Phi_\sigma^b )\right)
\right]
\det \left( g_{ab} \partial_c\Phi^b\sigma \right)\non\\
&=& \sum_{\Phi^a_\sigma =0} {\rm 
sign}\det \ || \partial_b\Phi^a_\sigma ||
\la{Z0}
\ea
In particular, if we select $s^a_\alpha =
\partial_\alpha\phi^a - \chi_\alpha^a$ and take $\chi_\alpha^a$ to be a
self--dual Hamiltonian vector field, {\it i.e.}
\be
\chi_\alpha^a ~ = ~ \half \partial^a H_\alpha(\tau,\phi)
\la{hvf}
\ee
where $H_\alpha(\phi)$ are two {\it a priori} arbitrary
Hamiltonian functions on
$X$ related by the self-duality condition 
for $\chi_\alpha^a$, we find that the integral localizes to
\be
\Phi_\sigma^a ~=~ \partial_\sigma\phi^a + {I^a}_b\partial_\tau\phi^b -
\partial^a H_\alpha(\tau,\phi)  ~=~ 0
\la{inflef}
\ee
which coincides with Floer's instanton equation (\ref{pseudo}). Thus
we have established that for this equation
\be
\sum_{\Phi^a_\sigma =0} {\rm sign}\det \ || \partial_b\Phi^a_\sigma ||
\ = \ \int [\d\phi^a][\d\psi^a] \Pfaff (\hat {R^a}_b)
\la{yhtalo}
\ee
This is a (formal) infinite dimensional analogue of the familiar Morse
theoretic relation between the Gauss-Bonnet-Chern and Poincar\'e-Hopf
formul\ae.  Note that demanding $\chi_\alpha^a$'s to be self--dual
together with (\ref{hvf}) implies that ${I^a}_b$ must be complex
structure so that $X$ is now a K\"ahler manifold.

In analogy with (\ref{lefs}), the present result can be viewed as an
infinite dimensional loop space version of the Lefschetz fixed point
theorem. For this, we remind that Floer's instanton equation
(\ref{pseudo}) is defined on the cylinder $S^1\times R$ with
coordinates $\tau\in [0,T]$ and $\sigma \in (-\infty, \infty)$.  For a
fixed $\sigma$ the $\tau$-dependence defines a closed loop on $X$, and
the $\sigma$-dependence parameterizes a smooth flow {\it i.e.} homotopy
mapping of these loops.  When $\sigma \to \pm \infty$ the bounded
orbits of (\ref{pseudo}) tend asymptotically to periodic solutions of
(\ref{trajectory}), implying that we have a flow between periodic
classical trajectories only.  The additional requirement of periodic
boundary condition $\phi(+\infty , \tau) = \phi(-\infty , \tau)$ then
identifies the periodic solutions of (\ref{trajectory}) with the fixed
points of the $\sigma$-flow in the loop space.  This generalizes the
setting of the conventional Lefschetz fixed point theorem that we have
discussed in connection of (\ref{lefs}) to the loop space. In
particular, if we specialize $H_\alpha$ in (\ref{yhtalo}) to coincide
with our original Hamiltonian we obtain a lower bound for the number
of periodic solutions to (\ref{trajectory}) in terms of the Witten
index of the topological $\sigma$-model,
\be
\#\{ \  T {\rm -periodic ~ trajectories} \ \}  ~ \geq ~  
\left| 
\int [\d\phi^a][\d\psi^a] \Pfaff (\hat {R^a}_b)
\right|
\la{viela}
\ee

The underlying idea in Floer's approach to the Arnold conjecture is
that the quantum cohomology of the topological $\sigma$-model should
coincide with the de Rham cohomology of the original symplectic
manifold {\it i.e.} the target manifold of the $\sigma$-model. Such a
relation would then provide a natural regularization of the infinite
dimensional Euler character in (\ref{yhtalo}), (\ref{viela}) and in
particular explains why an estimate such as (\ref{Arconj}) makes sense
as a Morse inequality. We shall now proceed to evaluate our path
integral using localization methods to establish that the Euler
character (\ref{euler}) of quantum cohomology indeed coincides with
the Euler character of the de Rham cohomology over the symplectic
manifold $X$.

For this, we specialize to a symplectic manifold which is K\"ahler.
We select local coordinates so that ${I^a}_b = i {\delta^a}_b$ and
${I^{\bar a}}_{\bar b} = -i{\delta^{\bar a}}_{\bar b}$. Self--duality
then implies that $F^{a}_z = F^{\bar a}_{\bar z} = 0$ so that the only
surviving components are $F^a_{\bar z} $ and $ F^{\bar a}_{z} $, and
similarly for $\rho^a_\alpha$.  Using the (formal) invariance of
(\ref{path}) under local variations of $\theta$, we introduce the
functional
\be
\theta ~=~ \eta^{\alpha\beta} g_{ab}
F^a_\alpha \rho^b_\beta ~+~ \mu
g_{ab} \eta^{\alpha \beta} \partial_{\alpha}
\phi^a \rho_{\beta}^b
\la{theta-2}
\ee
and consider the pertinent action (\ref{action}). Explicitly
(we set $F^a_{\bar z} \equiv F^a$ {\it etc}.),
\ba
S \ &=&  \ g_{a\bar b} F^a F^{\bar b} + {R^a}_{bc\bar d}
\psi^c \psi^{\bar d} \rho^b
g_{a\bar e} \rho^{\bar e} + {R^{\bar a}}_{\bar b c
\bar d} \psi^c \psi^{\bar d} \rho^{\bar
b} g_{\bar a e} \rho^e +
\mu ( F^a g_{a\bar b} \partial_{\bar z} \phi^{\bar b} +
F^{\bar a} g_{\bar a b} \partial_z \phi^b )
\non\\
&+&  \mu
\rho^a( - i g_{a \bar b} \partial_{\bar z} 
- i g_{a \bar d} \partial_{\bar z}
\phi^{\bar e} \Gamma^{\bar d}_{\bar b \bar e} ) \psi^{\bar b}
+ \mu \rho^{\bar a}( - i g_{\bar a b} \partial_{z} - i g_{\bar a
d}  \partial_{z}
\phi^{e} \Gamma^{d}_{b  e} ) \psi^{ b}
\la{action-2}
\ea
We evaluate the corresponding path integral in the $\mu
\to \infty$ limit, by separating the $z, \ \bar 
z$ independent constant modes
(for example in a Fourier decomposition) and scale 
the non-constant modes by
$\frac{1}{\sqrt{\mu}}$, {\it e.g.}
\[ 
\phi^a (z, \bar z)  \ = \ \phi_o^a  + {\hat
\phi}^a(z, \bar z)  \ \rightarrow \ \phi_o^a  
+ \frac{1}{\sqrt{\mu}} {\hat
\phi}^a(z, \bar z)
\]
and similarly for the other fields. Supersymmetry ensures that
the Jacobian for
this change of variables in (\ref{path}) is
trivial, and evaluating the integrals in
the $\mu \to \infty$ limit using the $\zeta$-function regularization we end
up with the Euler character of the phase space
$X$ in the form
\be
Z = \int d \phi_o^a d\phi_o^{\bar a} d \psi_o^a d \psi_o^{\bar a}
\ {\rm Pfaff} ( {R^a}_{b c \bar d} \psi_o^c \psi_o^{\bar 
d} ) {\rm Pfaff} (
{R^{\bar a}}_{\bar b c \bar d} \psi_o^c \psi_o^{\bar d} )
\la{finiteuler}
\ee
which also exhibits the underlying complex structure on $X$. As a
consequence, we have shown that the Euler characters in quantum
cohomology and de Rham cohomology {\it coincide} and
Floer's instanton equation defined over our torus obeys
\[ 
\sum_{\Phi^a_\sigma =0} {\rm sign}\det \ || 
\partial_c\Phi^b_\sigma || ~=~ \sum_k (-)^k B_k 
\] 
with $B_k$ the Betti numbers of the
symplectic manifold $X$. Obviously this is fully consistent with
(\ref{Arconj}) and relates 
our infinite dimensional version of
the Lefschetz fixed point formula for Floer's equation (\ref{yhtalo})
directly with its finite dimensional counterpart.

In conclusion, we have studied three {\it a priori} different
cohomologies: Floer's cohomology which describes periodic solutions to
Hamilton's equations, Witten's quantum cohomology which describes the
quantum ground state structure of a topological nonlinear
$\sigma$-model, and standard finite dimensional de Rham cohomology. By
investigating an infinite dimensional generalization of the familiar
Poincar\'e--Hopf and Gauss--Bonnet--Chern formul\ae$~$ together with
the Lefschetz fixed point theorem, we have shown that these three
cohomologies are intimately related. This is consistent with
the Arnold conjecture. In particular, our results indicate that
topological field theories and functional localization methods appear
to be useful tools also in the study of classical dynamical
systems.


\end{document}